\documentclass[useAMS,usenatbib,usegraphicx]{mn2e}

\usepackage{amssymb}

\title[Magnification of Type Ia Supernovae]{Revisiting the 
Magnification of Type Ia Supernovae with SDSS}

\author[M\'enard \& Dalal]{Brice M\'enard and 
  Neal Dalal\thanks{Hubble Fellow}\\
Institute for Advanced Study, Einstein Drive, Princeton NJ 08540 USA}

\begin{document}

\pagerange{\pageref{firstpage}--\pageref{lastpage}} \pubyear{2004}

\maketitle

\label{firstpage}

\begin{abstract}
We cross-correlate the sample of Type Ia supernovae from
\citet{riess04} with the SDSS DR2 photometric galaxy catalogue.  In
contrast to recent work, we find no detectable correlation between
supernova magnitude and galaxy overdensity on scales ranging between
$1-10\arcmin$.  Our results are in accord with theoretical
expectations for gravitational lensing of supernovae by large-scale
structure.  Future supernova surveys like SNAP will be capable of 
detecting unambiguously the predicted lensing signal.
\end{abstract}

\begin{keywords}
\end{keywords}

\section{Introduction}
Weak gravitational lensing by large-scale structure can both distort
and magnify the images of background sources.  The former effect,
cosmic shear, has been detected (see \citealt{refregier03} for a review)
and holds great promise for the study of matter fluctuations on large
scales.  The latter effect, cosmic magnification, has yet to be
detected convincingly but has been theoretically investigated in
detail \citep{jain03,menard03,takada03}.  In
particular, a number of authors have studied magnification of
Type Ia supernova, which have been employed as standard candles to
measure the expansion history of the universe
\citep[e.g.][]{tonry03,knop03}.  The RMS 
magnification  of supernovae is expected to be at the $\sim 1\%$
level on arcminute scales, with the largest magnifications
occurring near the centers of massive halos.

On scales larger than arcminutes,
only modest shear and magnification at the
sub-percent level are expected.  The theoretically expected magnitude
of shear fluctuations has been confirmed observationally in cosmic
shear surveys.  Accordingly, it would be extremely surprising if
observations contradicted the expected level of cosmic magnification,
since statistical shear and magnification are related in a simple way.
In principle, standard candles like Type Ia supernovae should offer a
means of measuring the magnification caused by large scale structure,
for example by cross-correlating supernova brightness with foreground
galaxy overdensity.  In practice, all but the very highest redshift
supernovae (which suffer the most lensing) are not expected to provide
useful measurements of cosmic magnification, since even Type Ia SNe
are not perfect standard candles.  The observed dispersion in
supernova magnitudes is roughly $\delta m\sim 20\%$, whereas
fluctuations of only $\lesssim 1\%$ are expected from lensing
\citep{menard03,takada03}.  Even if magnification
and galaxy overdensity were perfectly correlated, we would expect to
see only a weak correlation between supernova brightness and galaxy
overdensity, $r=\sqrt{\langle\delta\mu^2\rangle/\langle\delta
m^2\rangle}\lesssim0.05$.  Therefore a very large sample of SNIa would
be needed to observe such an effect.

Recently, \citet{williams04} have reported a highly significant
($>99\%$ confidence) detection of correlation between the magnitude of
55 supernovae and foreground galaxy overdensity on $10\arcmin$ scales,
corresponding to magnifications at the $\sim 10\%$ level.
\citet{wang04} has claimed further evidence for gravitational
magnification of the \citeauthor{riess04} supernovae.  If confirmed,
these results would confound much of the conventional wisdom on
magnification effects and would call into question previous cosmic
shear measurements which detected much lower shear variance on these
scales.  Additionally, if such a correlation exists, it is possible
to correct for it and therefore significantly reduce the dispersion in
supernova luminosities: the magnitude variance can be reduced by a
factor $1-r^2$ \citep{dalal03}.  This would be especially important
for modern supernova surveys, which can calibrate supernova magnitudes
nearly to 0.1 mag.

In this paper, we revisit this issue by cross-correlating the most
recent sample of cosmological Type Ia SNe compiled by \citet{riess04},
with low redshift galaxies in the Sloan Digital Sky Survey (SDSS)
photometric catalogue.  The plan of the paper is as follows.  In
section~\ref{sec:mag}, we briefly review theoretical predictions for
cosmic magnification on small scales.  Next, we describe the supernova
sample and galaxy catalogue employed in our analysis.  Results are
presented in \S\ref{sec:analysis}, and our conclusions are discussed in
\S\ref{sec:conclude}.

\section{Cosmic magnification} \label{sec:mag}

The large scale distribution of matter in the Universe can give rise
to measurable gravitational lensing effects (shear and magnification)
on background objects. The magnification factor $\mu$ depends on
whether matter along the line of sight is preferentially over- or
underdense compared to the mean.

The local properties of the gravitational lens mapping are
characterised by the convergence $\kappa$, which is proportional to
the surface mass density projected along the line-of-sight, and the
shear $\gamma$, which is related to the gravitational tidal field of
the lensing mass distribution.  The SDSS population of galaxies with
$m_r<21$ has a median redshift near 0.3 \citep[e.g.][]{dodelson02},
so that angular scales of arcminutes correspond to length scales of Mpc.
On these scales, the convergence and shear
fluctuations are small, $|\kappa|,|\gamma|\ll 1$.  To leading
order the magnification is approximately
\begin{equation}
  \mu=1+2\kappa+\mathcal{O}(\kappa^2,|\gamma|^2)\;,
\label{mu_exp}
\end{equation}
see \citet{menard03} for higher order corrections.
Observable effects are due to departures from the mean value of the
magnification. It is therefore convenient to use the overmagnification
$\delta\mu=\mu-1$. Then, the cross-correlation between $\delta\mu$ and
the foreground matter overdensity traced by galaxies at an angular
separation $\theta$ reads:
\begin{eqnarray}
  \langle\delta\mu\,\delta_\mathrm{gal}\rangle(\theta) &\approx&
2\;\langle\kappa\times\delta_\mathrm{DM}\rangle(\theta)\nonumber\\
&\approx& \frac{3}{2}\,\Omega_0\left(\frac{H_0}{c}\right)^2 \nonumber
  \int\frac{\mathrm{d} w}{a(w)}\,
\frac{D_d\;D_{ds}}{D_s}\,W_\mathrm{gal}(w)\nonumber\\
&&\times b_{\mathrm{gal}}\,
\int\frac{s\mathrm{d} s}{2\pi}P_\delta\left(\frac{s}{f_k(w)},w\right)
  \mathrm{J}_0(s\,\theta)\;.
\label{eq_crosscorrelation}
\end{eqnarray}
In
this expression, $w$ is the comoving distance along the line-of-sight,
$D_{d,s,ds}$ are angular diameter distances to lens, source, and from
lens to source respectively, $W_\mathrm{gal}$ is the normalised
distribution of galaxies along the line-of-sight, $a$ is the scale
factor and $P_\delta$ is the matter power spectrum, calculated using
$\Gamma=0.19$ and the fitting function of \citet{peacockdodds},
and we assume that the linear bias $b_{\mathrm{gal}}=1$ for the
galaxies and scales of interest \citep{verde02}.  In principle, the
bias of the detected galaxies should vary with redshift, but because
the galaxy distribution is dominated by a narrow redshift
interval, the use of a constant bias seems reasonable. 
If instead of the
two-point correlation, we wanted the cross-correlation with galaxy
overdensity smoothed on scale $\theta$ we would replace 
$\mathrm J_0(s\theta)$ by $2\mathrm{J_1}(s\theta)/s\theta$.

\begin{figure}
\begin{center}
\includegraphics[width=0.44\textwidth,height=.27\textheight]{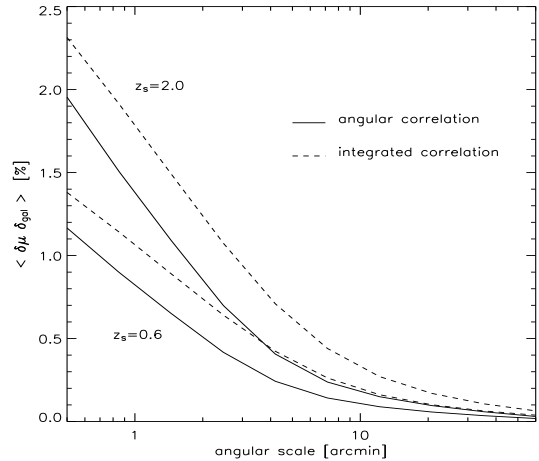}
\end{center}
\caption{Cross-correlation between gravitational magnification and
galaxy overdensity as a function of angular scale. A linear bias of
unity has been used in this calculation.}
\label{mag}
\end{figure}
Figure \ref{mag} shows the amplitude of this cross-correlation between
$\delta\mu$ and $\delta_\mathrm{gal}$ (Eq. \ref{eq_crosscorrelation}).
The solid line shows $\langle\delta\mu\,\delta_\mathrm{gal}\rangle$
computed for different angular separations and the dashed line
shows the integrated correlation, i.e.\ the correlation between
supernova magnitude  and the number of galaxies within radius
$\theta$.  For estimating these quantities we have used a galaxy redshift
distribution peaking at $z=0.3$ and a WMAP cosmology with
$\Omega_m=0.27$, $\Omega_\Lambda=0.73$, $n=1$, and $\sigma_8=0.9$
\citep{spergel03}. Black lines have been computed for supernova
redshifts of $z_s=0.6$ (the median redshift of our sample) and gray
lines for $z_s=2$. As can be seen in this figure, magnifications of
$\lesssim 1\%$ can be generated by structures on scales of a few
arcminutes or larger.

As an aside, it is perhaps worth noting that high redshift sources are
expected to be more strongly perturbed by lensing fluctuations, both
because the lensing efficiency increases with redshift, and because
more distant sources are subject to lensing from structures along a
greater path length.  For example, sources at $z=2$ are typically
magnified by $\delta\mu\sim 2\%$ by structures on scales of
1\arcmin.  This has ramifications for pencil-beam surveys
attempting to constrain dark energy with supernovae at high redshift,
$z\sim 1.5$.  For example, \citet{riess04} report the discovery of 16
SNIa in the GOODS field, over an area of 0.1 sq.\ degrees.
Lensing fluctuations are significantly correlated across this narrow
field.  If these high redshift supernovae played an important role in
the constraints derived by \citeauthor{riess04}, then their derived dark
energy parameters could be significantly biased, for example by 5\% in
$w$.  Fortunately, the 16 high redshift GOODS supernovae were
negligible within the larger sample of 177 SNe reported by
\citeauthor{riess04}, implying that their dark energy constraints were
not appreciably biased.  Future wide area surveys, like the SNAP
survey which expects to cover 20 sq.\ degrees, should also be
undisturbed by correlated structure across their fields.

\section{The data} \label{sec:data}

We attempt to detect a correlation between supernova magnitude and
galaxy overdensity using  following data sets:
\begin{itemize}
\item For supernovae, we use the ``gold'' sample of 156 SNIa compiled by
\citet{riess04}. For each supernova, the redshift, (dust-corrected)
distance modulus, and extinction correction have been provided by 
these authors. The SN redshifts range from $z=0.014$ to 1.75 with an
average value of $\approx 0.6$. 
\item For foreground galaxies, we use the second data
release (DR2) of the Sloan Digital Sky Survey (SDSS). As shown by
\citet{scranton02}  this dataset provides an efficient star/galaxy
separation for objects with $r<21$ and a homogeneous density by
requiring a seeing value greater than 1.5\arcsec.
\end{itemize}
Of the 156 SNe listed by \citeauthor{riess04}, 55 fall within the SDSS
DR2 footprint. Given the magnitude cut mentioned above, the average
galaxy density is roughly $1$ arcmin$^{-2}$.  The list of SNIa that we
use is given in table~\ref{tab:data}.

\begin{figure*}
\centerline{
\includegraphics[width=0.44\textwidth]{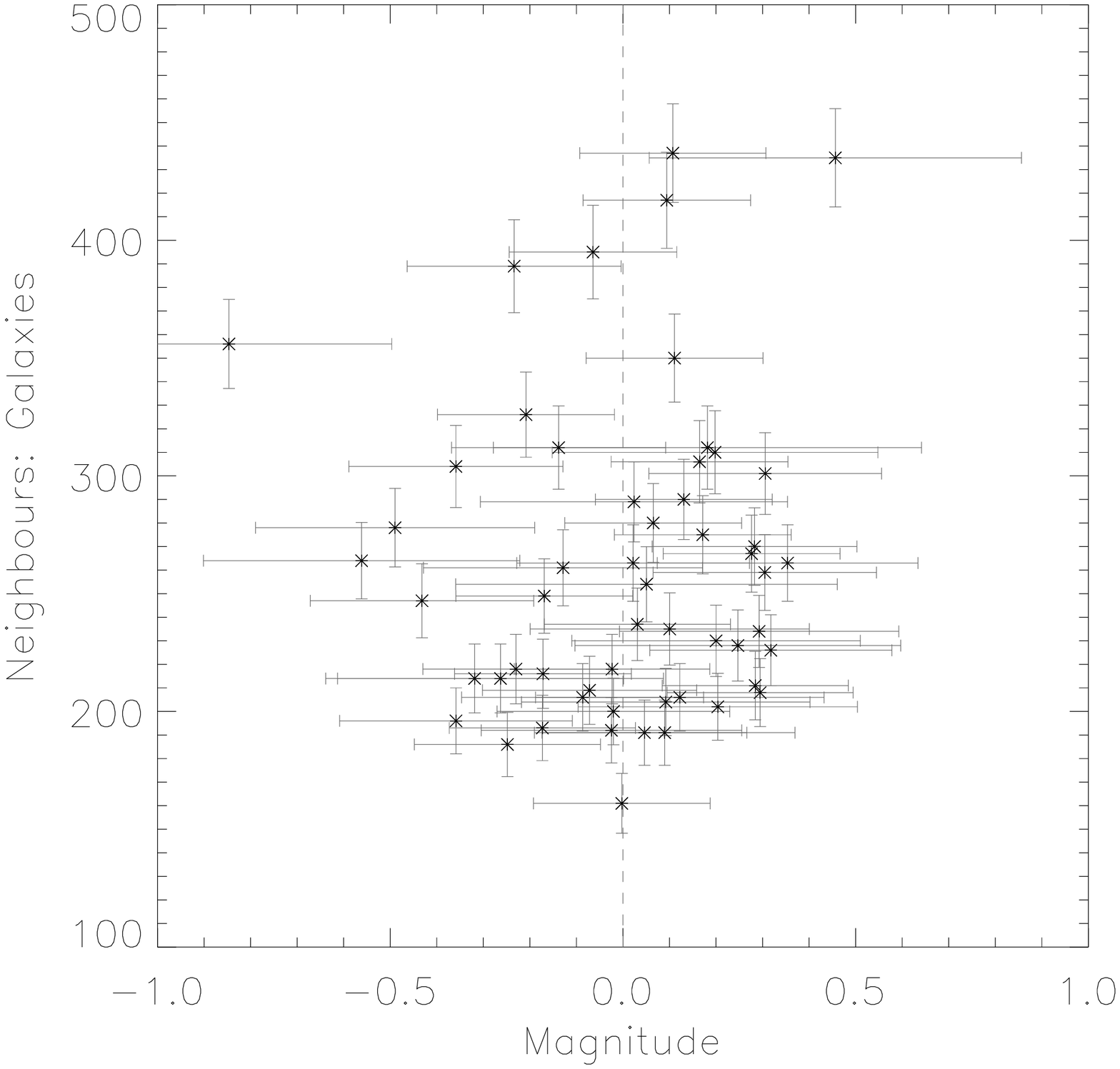}
\includegraphics[width=0.44\textwidth]{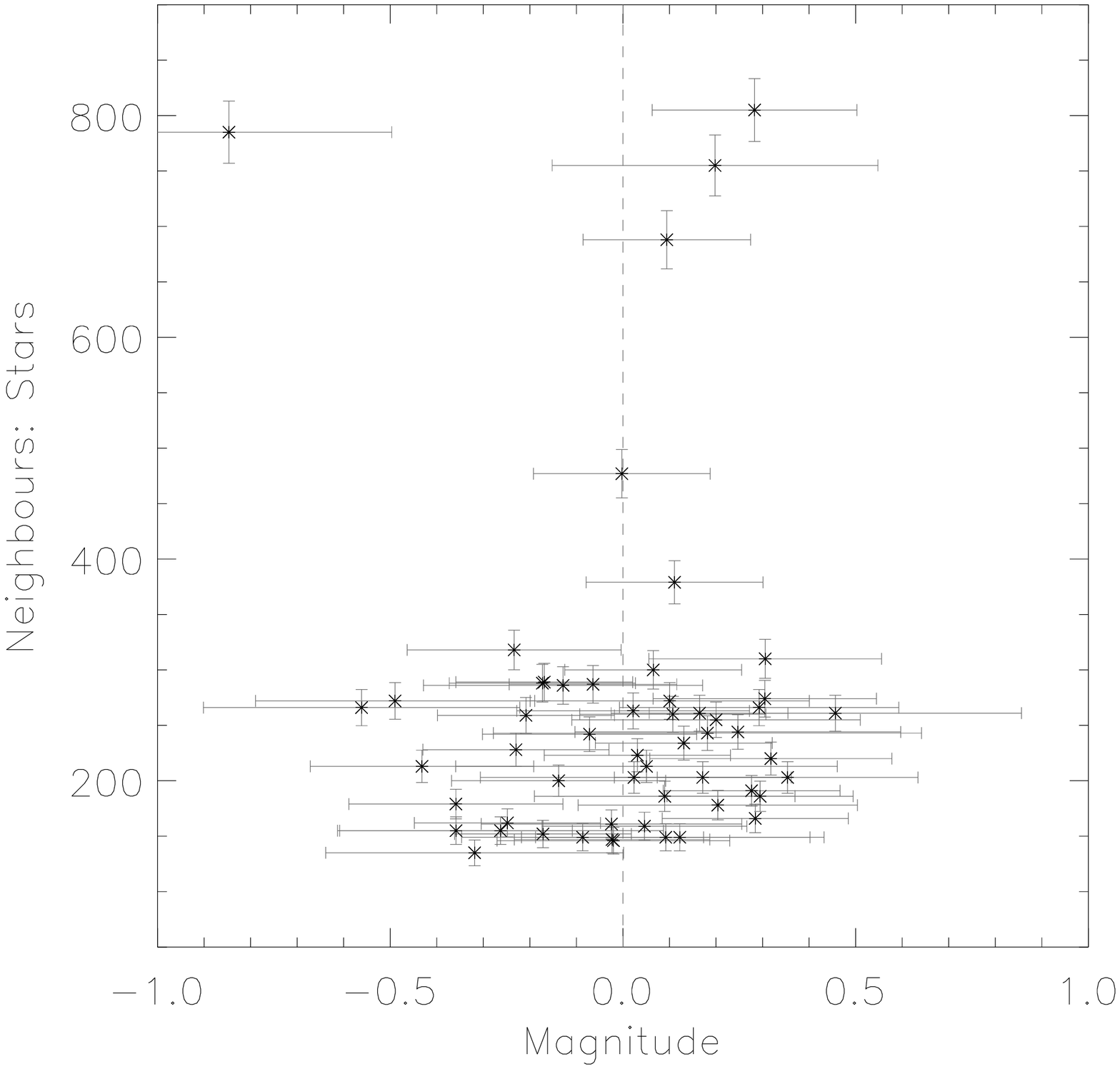}
}
\caption{Scatter plot of residual magnitude vs. number of galaxies
  (left) or stars (right) within 10\arcmin.
\label{scatterplot}}
\end{figure*}

\citeauthor{williams04} used magnitude cuts in each field in order to
restrict their galaxy counts to low redshifts objects and maximize the
signal. In our case, the bulk of SDSS photometric galaxies are at
redshifts $z\lesssim0.3$, while the median SN redshift in our sample
is $z_{\rm med}=0.6$.  Accordingly, the number of galaxies that will
overlap the supernova redshift distribution of our sample will be
largely negligible.



\section{Analysis} \label{sec:analysis}

For each of the 55 SNIa we first count the number of neighbours within an
angular radius $\theta$. The correlation between this number of
neighbours and the absolute SNIa magnitude is shown in
Fig.~\ref{scatterplot}. The left panel shows, as a function of
supernova magnitude, the distribution of galaxies located within
10\arcmin\ from the SNIa position, i.e. the angular scale where
\citet{williams04} found the strongest correlation.  The horizontal
error bars correspond to the measurement error in the SNIa magnitudes
quoted by \citet{riess04}, and the vertical errors correspond to the
Poisson noise of the galaxy counts. There is no apparent correlation
in this scatter plot; this will be quantified below. The right panel
shows the same correlation but using stars instead of galaxies and
allows us to check for possible systematics.

For each angular bin, we have computed the cross-correlation coefficient
\begin{equation}
r=\frac{\left\langle (\Delta N - \bar N)\;(\Delta M - \bar M) \right\rangle}
{\sqrt{\langle (\Delta N - \bar N)^2 \rangle
\langle (\Delta M - \bar M)^2 \rangle}}\,,
\end{equation}
where $N$ is the number of neighbours (galaxies or stars) within a
given angular separation from the supernova and $M$ is the SN absolute
magnitude. Neglecting the
measurement uncertainties in each object, and accounting only for the
scatter in the population, the error in the cross-correlation
coefficient reads
\begin{equation}
\sigma_r=\frac{1-r^2}{\sqrt{N_\mathrm{SN}}}\,,
\end{equation}
where $N_\mathrm{SN}$ is the number of SNIa. For cells of 10 \arcmin
radius, we find $r=0.03 \pm 0.13$ and $r=-0.04 \pm 0.26$ for galaxies
and stars respectively. The result obtained by \citet{williams04} is
therefore not reproduced in our analysis: no significant correlation
is detected.

\begin{figure}
\centerline{
\includegraphics[width=0.44\textwidth]{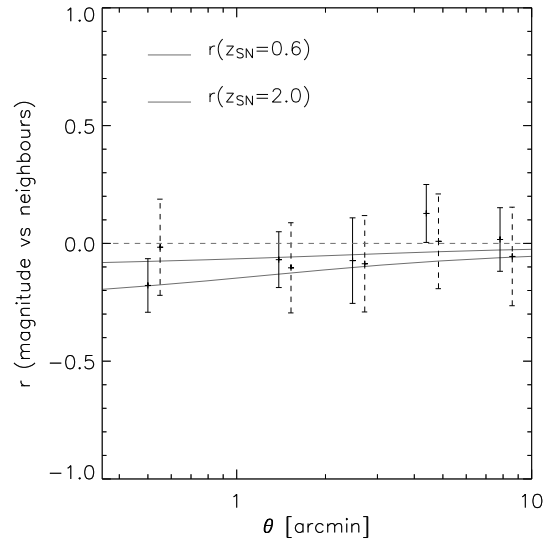} 
}
\caption{Cross-correlation coefficient $r$ between SN magnitude and
  galaxy counts (solid) or star counts (dashed), as a function of
  radius.  The overplotted curves show the theoretical prediction for
  source redshifts of 0.6 and 2.}
\label{r_vs_theta}
\end{figure}

In Fig.~\ref{r_vs_theta} we plot the cross-correlation coefficient $r$
as a function of $\theta$.  In this figure, we have counted galaxies
within an annulus centred on each $\theta$, so that the plotted errors
are uncorrelated.  For both galaxies and stars, no significant
correlation has been detected on any scale.


\section{discussion} \label{sec:conclude}

Using the ``gold'' sample of type Ia supernovae given by
\citet{riess04} and the galaxy distribution measured by SDSS DR2, we
have investigated correlations between the brightness of the
supernovae and the density of foreground galaxies, on scales of
1-10\arcmin. Our analysis does not reproduce the recent results of
\citet{williams04}, who detected a cross-correlation at high
significance ($>99\%$ confidence) on 10\arcmin\ scales.  Our result is
in line with theoretical predictions: the $\lesssim1\%$ rms
magnification caused by large-scale structure on arcminute scales is
negligible compared to the observed $\sim 20\%$ dispersion in
supernova magnitudes.

The number of SNe used by \citeauthor{williams04} closely matches the
number used in our analysis.  The main difference between the two
analyses is that \citeauthor{williams04} derive galaxy counts from the
APM catalogue, whereas we have used counts from the SDSS photometric
catalogue.  While the APM catalogue is based on scans of photographic
plates taken with the UK Schmidt telescope, the SDSS uses modern CCDs
with high-quality imaging reaching down to much deeper magnitudes,
providing a higher galaxy density.  Therefore we expect the SDSS to be
more suitable for investigating weak correlations.

Thus there is no evidence for a significant correlation between
supernova brightness and foreground galaxy overdensity.  Such a
correlation would be expected for a sample of weakly lensed
supernovae.  
In this light, the recently claimed detection of significant weak
lensing in the \citet{riess04} sample by \citet{wang04} is somewhat
surprising.  
Further examination of the data presented by \citeauthor{wang04}
indicates that the effect appears to be dominated by a few
outliers in the magnitude distribution, which \citeauthor{wang04}
ascribes to magnification factors $\sim2$, i.e.\ strong lensing.  This
would imply a surprisingly high strong lensing rate ($\sim5\%$) among
high $z$ supernovae, inconsistent with the strong lensing rate ($\sim
10^{-3}$) of distant quasars \citep{class}.

Lastly, we note that the theoretically expected SN-galaxy correlation
(solid curves in figure~\ref{r_vs_theta}) would be
detectable with larger samples of supernovae with smaller flux errors.
Although the signal is not currently detectable using the
\citet{riess04} supernovae and SDSS galaxies, future datasets
(e.g.\ SNAP)
should unambiguously exhibit the lensing signal on small scales.  

\section*{Acknowledgments}
BM is supported by a fellowship of the F. Gould foundation. ND is supported by
NASA through Hubble Fellowship grant \#HST-HF-01148.01-A, awarded by
the Space Telescope Science Institute, which is operated by the
Association of Universities for Research in Astronomy, Inc., for NASA,
under contract NAS 5-26555.

\appendix

\begin{table*}
\caption{Supernovae used in our analysis.  Here, $\mu_0$ and $\sigma$
  are the distance modulus and its error.  All quantities are taken
  from the larger ``gold'' sample of \citet{riess04}.
\label{tab:data}}
\begin{tabular}{lllcccc}
\hline
~~SN    & ~~RA  & ~~DEC  & redshift & $\mu_0$ & $\sigma$ & host $A_v$ \\
\hline\hline
1999dk  &01 31.5  &+14 17  &0.0141  & 34.43  & 0.26  & 0.20       \\
1993ae  &01 29.8  &-01 59  &0.0180  & 34.29  & 0.23  & 0.00       \\
1994M  & 12 31.2  &+00 36  &0.0244  & 35.09  & 0.20  & 0.23       \\
1994Q  & 16 49.9  &+40 26  &0.0290  & 35.70  & 0.19  & 0.33       \\
1998cs  &16 30.7  &+41 13  &0.0327  & 36.08  & 0.19  & -0.03       \\
1994T  & 13 19.0  &-02 09  &0.0360  & 36.01  & 0.20  & 0.09       \\
1997I  & 04 59.6  &-03 09  &0.172  &  39.79  & 0.18  & ---       \\
1999fw  &23 31.9  &+00 10  &0.278  &  41.00  & 0.41  & 0.26       \\
1997bj  &10 42.4  &+00 02  &0.334  &  40.92  & 0.30  & 0.34       \\
1996K  & 08 24.7  &-00 21  &0.380  &  42.02  & 0.22  & 0.02       \\
1995ba  &08 19.1  &+07 43  &0.3880  & 42.07  & 0.19  & ---       \\
1997am  &10 57.5  &-03 14  &0.416  &  42.10  & 0.19  & 0.00       \\
1997bh  &13 44.6  &-00 20  &0.420  &  41.76  & 0.23  & 0.60       \\
1997Q  & 10 56.9  &-03 59  &0.430  &  41.99  & 0.18  & ---       \\
1998ba  &13 43.6  &+02 20  &0.430  &  42.36  & 0.25  & ---       \\
1997aw  &10 23.5  &+04 07  &0.440  &  42.57  & 0.40  & 0.80       \\
1997ai  &10 49.0  &+00 32  &0.450  &  42.10  & 0.23  & ---       \\
1999ff  &02 33.9  &+00 33  &0.455  &  42.29  & 0.28  & 0.19       \\
1997P  & 10 55.9  &-03 57  &0.472  &  42.46  & 0.19  & ---       \\
2002dc  &12 36.8  &+62 13  &0.475  &  42.14  & 0.19  & 0.23       \\
1995ay  &03 01.1  &+00 21  &0.480  &  42.37  & 0.20  & ---       \\
1996ci  &13 45.9  &+02 27  &0.495  &  42.25  & 0.19  & ---       \\
1997cj  &12 37.1  &+62 26  &0.500  &  42.74  & 0.20  & 0.15       \\
1999U  & 09 26.7  &-05 38  &0.500  &  42.75  & 0.19  & 0.04       \\
1997as  &08 24.2  &-00 48  &0.508  &  41.64  & 0.35  & 0.85       \\
1997bb  &12 29.0  &+00 09  &0.518  &  42.83  & 0.30  & 0.11       \\
2001iy  &10 52.4  &+57 17  &0.570  &  42.88  & 0.31  & -0.04       \\
1996I  & 12 00.7  &-00 16  &0.570  &  42.81  & 0.25  & 0.14       \\
1997af  &08 23.9  &+04 09  &0.579  &  42.86  & 0.19  & ---       \\
1995ax  &02 26.4  &+00 49  &0.615  &  42.85  & 0.23  & ---       \\
1996H  & 12 28.9  &-00 05  &0.620  &  43.11  & 0.30  & 0.09       \\
1998M  & 11 33.7  &+04 05  &0.630  &  42.62  & 0.24  & 0.75       \\
2003be  &12 36.4  &+62 07  &0.64  &   43.07  & 0.21  & 0.23       \\
1997R  & 10 57.3  &-03 55  &0.657  &  43.27  & 0.20  & ---       \\
2003bd  &12 37.4  &+62 13  &0.67  &   43.19  & 0.28  & 0.27       \\
2001ix  &10 52.3  &+57 07  &0.710  &  43.05  & 0.32  & 0.53       \\
1998bi  &13 47.7  &+02 21  &0.740  &  43.35  & 0.30  & ---       \\
1997ez  &08 21.6  &+03 25  &0.778  &  43.81  & 0.35  & ---       \\
2001hx  &08 49.4  &+44 02  &0.798  &  43.88  & 0.31  & 0.31       \\
2001hy  &08 49.8  &+44 15  &0.811  &  43.97  & 0.35  & 0.03       \\
1999fj  &02 28.4  &+00 39  &0.815  &  43.76  & 0.33  & 0.23       \\
2001jf  &02 28.1  &+00 27  &0.815  &  44.09  & 0.28  & 0.23       \\
1996cl  &10 57.0  &-03 38  &0.828  &  43.96  & 0.46  & ---       \\
1997ap  &13 47.2  &+02 24  &0.830  &  43.85  & 0.19  & ---       \\
2003eq  &12 37.8  &+62 14  &0.839  &  43.86  & 0.22  & 0.22       \\
2001jh  &02 29.0  &+00 21  &0.884  &  44.23  & 0.19  & -0.01       \\
2003eb  &12 37.3  &+62 14  &0.899  &  43.64  & 0.25  & 0.26       \\
2003lv  &12 37.5  &+62 11  &0.94  &   43.87  & 0.20  & 0.15       \\
2002dd  &12 36.9  &+62 13  &0.95  &   44.06  & 0.26  & 0.24       \\
2003es  &12 36.9  &+62 13  &0.954  &  44.28  & 0.31  & 0.07       \\
2002ki  &12 37.5  &+62 21  &1.140  &  44.84  & 0.30  & 0.09       \\
1999fv  &23 30.6  &+00 17  &1.19  &   44.19  & 0.34  & 0.24       \\
2003az  &12 37.3  &+62 19  &1.265  &  45.20  & 0.20  & 0.25       \\
2003dy  &12 37.2  &+62 11  &1.340  &  45.05  & 0.25  & 0.54       \\
1997ff  &12 36.7  &+62 13  &1.755  &  45.53  & 0.35  & 0.00       \\
\hline
\end{tabular}
\end{table*}


\newpage

\label{lastpage}

\end{document}